# Ultrahigh-Speed Rotating Nanoelectromechanical System (NEMS) Devices Assembled from Nanoscale Building Blocks

Kwanoh Kim[1*], Xiaobin Xu[2*], and D. L. Fan[1, 2]

[1] Department of Mechanical Engineering, the University of Texas at Austin, Austin, TX 78712, USA
[2] Materials Science and Engineering Program, the University of Texas at Austin, Austin, TX 78712, USA
Correspondence can be addressed to: dfan@austin.utexas.edu

*equal contribution

**Abstract**

The development of rotary nanomotors is crucial for further advancing NEMS technology. In this work, we report innovative design, assembly, and rotation of ordered arrays of nanomotors. The nanomotors were bottom-up assembled from nanoscale building blocks with nanowires as rotors, patterned nanomagnets as bearings, and quadrupole microelectrodes as stators. Arrays of nanomotors can be synchronously rotated with controlled angle, speed (to at least 18,000 rpm), and chirality by electric fields. Using analytical modeling, we revealed the fundamental nanoscale electrical, mechanical, and magnetic interactions in the nanomotor system, which excellently agrees with experimental results and provides critical understanding for designing future metallic NEMS devices. The nanowire motors were applied for controlled biomolecule release, and for the first time demonstrated releasing rate of biochemicals on nanoparticles can be tuned by mechanical rotation. The innovations reported in this research, from concept, design, actuation, to application are relevant to NEMS, nanomedicine, microfluidics, and lab-on-chip architectures.

Nanoelectromechanical System (NEMS) devices, consisting of both electronic and mechanical components are emerging as the next-generation technology that can significantly impact people's lives. It has intrigued the research community for over a decade, not only due to the rich fundamental science where devices are made on the nanoscale,[1] but also due to the high potential of making technical breakthroughs in various areas including robotics,[2] biomedical research,[3-6] and optoelectronics.[7]

Rotary nanomotors, a type of NEMS devices, are particularly critical for advancing NEMS technology.[8,9] Traditional fabrication of miniature motors requires complex design and arduous processes that have greatly hindered the development of rotary NEMS.[10-13] For instance, a simple rotary MEMS device requires fabrication and integration of multiple components including rotors, bearings, hubs, and stators all on miniaturized scales.[9-11,14] Using traditional MEMS technologies— photolithography and micromachining borrowed from the microelectronics industry, hundreds of fabrication steps are required. Such complexity has restricted the size of the motors to be in the range of millimeters to hundreds of micrometers.[15-17] Very few can make truly nanoscale motors even using the best available techniques.[11,13] Moreover, the devices suffer from low yield and short lifetime.[18] Only several rotary MEMS were made on each wafer and they operated for just a few hours.[10,15] It is highly desirable to investigate new mechanisms to realize large arrays of rotary NEMS devices with high efficiency, nanoscale dimensions, reliable performance, and at a low cost.

In this work, we report innovative design and mechanisms for assembling and operating arrays of rotary NEMS devices made from nanoscale building blocks. The nanomotors consist of multisegment nanowires, patterned nanomagnets, and quadrupole microelectrodes as rotors, bearings, and stators, respectively as shown in Fig. 1(a) and (b). Arrays of nanomotors can be assembled and synchronously rotated with controlled angle, speed (to at least 18,000 rpm), and chirality. The fundamental electric, magnetic, and mechanical interactions involved in the components of the nanomotor systems were investigated, which provide understanding for designing and actuating various metallic NEMS devices. The innovation in this research may inspire multiple research fields including MEMS/NEMS, bio-NEMS, micro/nanofluidics, and lab-on-a-chip architecture.

Recently, intensive research efforts were focused on using nanoentities as actuation components for MEMS/NEMS devices due to the unique advantages provided by nanotechnology: (1) large arrays of nanoentities with controlled geometry, chemistry, physical and mechanical properties can be routinely synthesized at low cost.[19] (2) The size of the MEMS devices can be significantly shrunk by using synthesized nanoentities as building blocks.[1] (3) The unique physical/chemical/electrical properties of nanoparticles improve the performance of miniaturized mechanical devices. [20,21]

A group at UC Berkeley fabricated nanomotors with top-down multi-step electron-beam lithography.[12] In such a device, a nanoscale metal pad is attached to a piece of suspended multiwall carbon nanotube (MWCNT) via E-beam lithography. With high electric voltages, the metal pad rotates around the MWCNT with the inner walls of carbon nanotubes serving as bearings. The excellent demonstration of nanoentities for rotary NEMS, however, requires complex fabrication procedure. Recently, a group at Cornell University made another type of nanomotors based on inorganic and organic hybrid structures.[13] Molecular motors, a type of rotary protein, were used as the driving component of the nanomotors. Lithographically patterned inorganic nanorods attached to the molecular motors were used as rotors. When hydrolyzing adenosine triphosphate (ATP), the molecular motors can make the inorganic nanorods rotate. Due to the complexity of the conjugation of inorganic nanorods to molecular motors, only a few nanorods out of hundreds had been rotated. Also, limited by the characteristics of natural molecular motors, the devices cannot alter either their rotation or speed. Catalytic nanomotors have emerged as a new type of nano-mechanical devices that convert chemical energy to mechanical motions.[8,22,23] Most of the catalytic motors have been actuated to transport biomolecules, such as proteins or bacteria.[24] Recently, it was demonstrated that asymmetrical and one-end fixed nanorods can rotate,[25,26] nevertheless, with random speed, orientation, and locations.

Here, we investigated a unique type of nanomotors made of strategically assembled nanoentities, such as nanowires and nanodisks, as shown in Fig. 1(b). The nanomotors consist of multisegment Au/Ni/Au nanowires (20–400 nm in diameters and 6–10 μm in length) acting as rotors, patterned nanomagnets (500 nm–2 μm in diameter) acting as bearings, and simple quadruple microelectrodes acting as stators. The patterned nanomagnets are the core components for the anchorage of the rotary parts. The nanomagnets consist of tri-layer thin-film stacks of Au/M/Cr, where M representing magnetic materials such as Ni and Co. Each layer in the nanomagnets serves a purpose: the bottom Cr layer adheres to the substrate, the middle magnetic layer provides the magnetic field, and the top Au layer is used for adjusting the spacing between the magnetic layer and the nanowires in order to tune the magnetic attraction. The magnetic attraction can be adjusted so precisely that it can anchor the nanowires to the bearings, but not too tightly to prevent the nanowires from rotating.

To realize such nanomotors, there are two fundamental problems to be addressed: (1) *how to precisely assemble nanomotors with nanowires anchoring on the patterned nanomagnets? (2) How to rotate the nanomotors?* Here, we exploited the use of "electric tweezers" to resolve these problems. Electric tweezers are our recently invented nano-manipulation technique, which can transport nanoentities, such as nanowires and nanotubes, along arbitrary trajectories to designated positions with a precision of 150 nm[27-29] and rotate them with prescribed angle, speed, and chirality.[30]

The nanowire rotors, vital components of the rotrary nanomotors, were designed with a three-segment Au/Ni/Au (4.5 µm/1 µm/4.5 µm, diameters 50 nm- 300 nm) structure, where the Ni segment serves to anchor on the patterned magnetic bearings. Here, Au/Ni/Au nanowires were fabricated by electrodeposition into nanoporous templates in a three-electrode cell setup as detailed in the experimental session.[27,31] By using electric tweezers, the nanowires can be transported by the DC *E* field and aligned either parallel (AC//DC) or perpendicular (AC⊥DC) to their transport directions by the AC *E* field. By controlling the durations of the combined AC and DC *E* fields in both the X and Y directions, the nanowires can be precisely manipulated in two dimensions along prescribed trajectories to the patterned magnetic bearings as shown in Fig. 1(c) and Video S1. When positioned in the vicinity of a nanobearing, the nanowire is swiftly attracted and assembled atop of the nanobearing by the magnetic force between the nanowire and nanobearing. Once anchored on the nanobearing, the nanowire cannot be moved by the *E* fields any more. In such a manner, we can readily form ordered arrays of nanomotors by transporting and assembling nanowires in series. Due to the efficient manipulation by the electric-tweezers, each assembling event only takes a few seconds.

The nanobearings (500 nm – 2 μm in diameter), anchoring nanowire rotors, consist of a thin film stack of Cr (6nm)/ Ni (80 nm)/Au (100 nm), where Cr adheres to the substrate, Ni with a gently tilted perpendicular magnetic anisotropy (PMA) provides magnetic anchoring force [support information S6],[32] and Au works as a separation layer to tune the magnetic field from Ni [Fig. 1(b)]. The in-plane magnetic orientation of the nanobearings is controlled by demagnetization with a gradually reducing oscillating magnetic field of ±10 kG followed by re-magnetization at 10 kG along a desired direction for 5 seconds.

Applying four AC voltages (10 kHz – 150 kHz, 8 – 17 V) with a sequential phase shift of 90° on the quadruple microelectrodes, we created a rotating AC $E$ field, which can instantly drive arrays of nanowires to rotate synchronously atop of the magnetic bearings. The motions of the rotating nanomotors were captured by an optical microscope equipped with a CCD camera operating at 14 – 65 frame/sec (fps). The nanomotors were rotated both clockwise (CW) and counterclockwise (CCW), same as that of the rotating $E$ fields as shown in Fig. 1(d). At each AC voltage, the rotation instantly started, reached a terminal velocity, and stopped as soon as the external $E$ field was removed without observable acceleration or deceleration [Fig. 1(e)]. The instant rotation response is due to the extremely low Reynolds number of $10^{-5}$ for nanowires in deionized (D.I.) water, where the viscous force overwhelms the motion.[33] Taking a closer look, we noticed that the plot of rotation angle ($\theta$) versus time ($t$) was not completely linear [Fig. 2(a)]. By conducting the first derivative, rotation speed ($\omega$) can be obtained, which showed a clear sinusoidal dependence on time ($t$) and angle ($\theta$), oscillating between two states: the high speed ($\omega_{Max}$) and low speed states ($\omega_{min}$) with a periodicity of 360° [Fig. 2(b) and (c)]. To understand these phenomena, a thorough investigation of electric, magnetic, and mechanical forces/torques in the nanoscale components of a nanomotor system is highly desirable.

Five torques can be identified in the nanomotor system given by:

$$\tau_e + \tau_{e'} + \tau_\eta + \tau_M + \tau_f = 0, \qquad (1)$$

where $\tau_e$, $\tau_{e'}$, $\tau_\eta$, $\tau_M$, and $\tau_f$ are the electric torque due to the rotating AC $E$ fields, effective torque due to the induced electric polarization of the nanowire and bearing, viscous drag torque from the water medium, magnetic and frictional torques between the magnetic segment in the nanowire and the magnetic nanobearing, respectively.

The viscous drag torque $\tau_\eta$ on a rotating nanowire in DI water is calculated as:[34]

$$\tau_\eta = \frac{1}{3}\omega\pi\eta l^3 \frac{N^3 - N}{N^3\left[\ln\left(\frac{l}{Nr}\right) + 0.5\right]} = -1.74 \times 10^{-19}\omega \text{ [N·m]}, \qquad (2)$$

where $r$ and $l$ are the radius and length of the nanowire, respectively, $\eta$ is the viscous coefficient of suspension medium, and $N$ is a constant, which is taken as 2. For a nanowire of 150 nm in radius and 10 μm in length, $\tau_\eta = -1.74 \times 10^{-19}\omega$ N·m. The electric torque is due to the interaction of the polarized nanoparticles with the AC $E$ fields, given by[35]

$$\tau_e = \frac{2\pi}{3} r^2 l \varepsilon_m \text{Im}(K) E^2 = aE^2 \qquad (3)$$

where $\varepsilon_m$ and Im($K$) are the permittivity of the suspension medium and the imaginary part of the Clausius-Mossotti factor $K$ of the nanowire, respectively. Since $\tau_e$ solely depends on the suspension medium, Im($K$), and the dimension of a nanowire, the value of $\tau_e$ for an assembled nanowire rotor should be identical with that on the same nanowire when it freely rotates in suspension. Note that for a free rotating nanowire, the electric torque balances with the viscous torque ($\tau_e = \tau_\eta$). Therefore, by combining equations (2) and (3), the electric torque ($\tau_e$) for the assembled nanowire rotor can be readily calculated from that of a free rotating nanowire, and the rotation speed $\omega \sim V^2$. Indeed, the linear dependence of $\omega$ versus $V^2$ was experimentally observed, with a slope of $\frac{a}{1.74 \times 10^{-19}} = 9.91$ deg/s·$V^2$ as shown in Fig. 3(a, orange line, for the free rotating nanowire). Therefore, the electric torque can be readily determined as $\tau_e = 3.01 \times 10^{-20} E^2$ N·m/V$^2$. At a given voltage, e.g. 12 V, the nanowire rotor receives an electric torque of 4.33 pN·μm.

The induced torque ($\tau_{e'}$) due to the electrically polarized nanorotor and nanobearing in an external AC $E$ field has a much lower, though non-negligible effect, on the nanomotor than that of the driving electric torque ($\tau_e$). This induced torque played two roles on the rotation of nanomotors, both of which are proportional to $E^2$ and thus $V^2$: first, it directly hindered the rotation of nanorotors; second, it resulted in an additional frictional torque between the rotor and bearing besides that arisen from the magnetic attraction. These two effects can be known from the slopes of $\omega$–$V^2$ plots in Fig. 3(a), where the rotation slopes of a nanowire motor (blue, black, and red) are consistently lower than that of the same nanowires when freely rotating (orange). Note that no other torques, either the magnetic ($\tau_M$) or frictional torques ($\tau_f$), given in Eq. (1) depends on the external $E$ fields. Therefore, the reduced slope is a sole result of the induced torques between the electrically polarized nanowire and bearing. Based on this understanding, we can calculate the induced torque as $\tau_{e'} = bE^2$ and obtain $b = -2.52 \times 10^{-21}$ N·m/V$^2$ from rotation of the free/anchored nanowire in Fig. 3(a). The $b$ value is approximately 1/10 of that for the electric torque ($\tau_e$), showing the non-negligible effect of the induced electric torques in components of metallic NEMS devices.

Given the drag ($\tau_\eta$), electric ($\tau_e$), and induced torques due to **E** field ($\tau_{e'}$), the sum of the non-**E**-field dependent magnetic ($\tau_M$) and frictional torques ($\tau_f$) exerted on the nanowire rotors can be readily known from Eq. (1). The combined torques of ($\tau_M + \tau_f$) exhibit a sinusoidal feature as shown in Fig. 2(d), which count for the angle-dependent periodic rotation of nanomotors in Fig. 2(b) and (c). To explicitly understand the contributions of the magnetic and frictional torques, we modeled the system using a simplified magnetic dipole-dipole interaction.[36] As shown in Fig.1(b), the angle dependent magnetic torque between the rotor and bearing is determined by the horizontal magnetic moments of the nanowire (***m₁***) and bearing (***m₂***), given by $\tau$(r, $\Theta$, *m₁*, *m₂*)= $\mu_0(m_1 m_2 \sin\Theta)/(4\pi x^3)$, and the angle-dependent magnetic force *is* F(r, $\Theta$, *m₁*, *m₂*)= $3\mu_0(m_1 m_2 \cos\Theta)/(4\pi x^4)+c$, where $\mu_0$ is the magnetic permittivity of vacuum, $\Theta$ is the angle between ***m₁*** and ***m₂***, x is the separation distance between the nanowire and the Ni layer in the nanobearing, and the constant $c$ is the magnetic force due to the magnetic moments of the nanobearing in the vertical direction (***m₃***) and the horizontal magnetic moment of the nanowire (***m₁***). If the friction coefficient at the interface of the nanowire and bearing is $\mu$, the frictional torque $\tau_f = \mu F$, then ($\tau_M + \tau_f$) can be written as:

$$\tau_M + \tau_f = d \sin(\theta - \theta_M) + e \cos(\theta - \theta_M) + f \quad (4)$$

Where $d$, $e$, and $f$ are constants, $\theta$ and $\theta_M$ are the angular positions of magnetic orientation of the nanowire and nanobearing, respectively, and $\Theta = \theta - \theta_M$. Fitting the experimentally obtained torque of ($\tau_M + \tau_f$) versus $\theta$ in Fig. 2(d), we readily extracted the angular dependent magnetic torque of $\tau_M = -8.19 \times 10^{-19} \sin(\theta - 27°)$ N·m and frictional torque of $\tau_f = -1.71 \times 10^{-19} \cos(\theta - 27°) - 5.35 \times 10^{-19}$ N·m at 12 V, 10 kHz as shown in Fig. 2(e). By using the same approach, the values of $d$, $e$, $f$, and $\theta_M$ were also determined for 10 and 11 V, which show excellent consistence with those obtained at 12 V (supporting information S7). These results provide great support for our modeling, where the magnetic and its resulting frictional torque are independent of the applied electric field. We also note that the highest rotation speed ($\omega_{MAX}$) occurred neither when the magnetic torque has the highest value ($\tau_{M,Max}$) in the same direction with the driving **E** field, when the magnetic torque equals to zero ($\tau_{M,0}$) aligning with the magnetic orientation of the bearing, nor when the frictional torque has the lowest value ($\tau_{f,min}$). The highest/lowest speed occurred when the sum of ($\tau_M + \tau_f$) was maximum/minimum in the orientation of the electric torque [Fig. 2(g) and inset of Fig. 2(f)].

The feasibility of the modeling and torque analysis can be further confirmed by plotting the experimentally obtained highest ($\omega_{MAX}$, blue) and lowest rotation speeds ($\omega_{MIN}$, red) of the nanomotors at each applied voltage as functions of $V^2$ [Fig. 3(a)]. Both $\omega_{MIN}$ and $\omega_{MAX}$ linearly increase with $V^2$, parallel to that of the average rotation speed ($\omega_{AVG}$, black) versus $V^2$, but vertically offset. Combining Eq. (1) – (4), we obtained

$$\omega = \frac{a+b}{1.74 \times 10^{-19}} E^2 + \frac{\sqrt{d^2 + e^2} \sin(\theta - \theta_M + \delta) + f}{1.74 \times 10^{-19}} \tag{5}$$

where $\tan \delta = e/d$. $\omega_{MAX}$ and $\omega_{MIN}$ occur when $\sin(\theta - \theta_M + \delta) = \pm 1$, respectively. The $E^2$ dependence in Eq. (5), accounts for the $V^2$ dependence of $\omega_{MIN}$ and $\omega_{MAX}$ in Fig.3(a). The constant slope $\frac{a+b}{1.74 \times 10^{-19}}$ of $\omega$–$E^2$, accounts for the parallel slopes of $\omega_{MAX}$–$V^2$, $\omega_{MIN}$–$V^2$, and $\omega_{AVG}$–$V^2$ in Fig. 3(a). Moreover, the x-intercept of $\omega_{MIN}$–$V^2$ can be used to estimate the threshold voltage ($V_{th}$) required to initiate the rotation of nanomotors. It predicted a threshold voltage value of $\sqrt{57.7} = 7.6$ V for a motor with a 2-µm-diameter bearing. This predicted value is very close to that of experimentally obtained 7.3 V, as the nanomotor failed to rotate a complete cycle up to 7.2 V [Fig. 3(b)]. It indicates that the minimum required voltage to initiate the nanomotor rotation should overcome the maximum value of the combined magnetic and frictional torques $(\tau_M + \tau_f)$ countering the electric torque. In summary, our modeling qualitatively revealed the distinct roles of various nanoscale torques involved in the nanomotor system, which excellently agreed with the experimental results. The new understandings from this investigation can be applied to design of various metallic NEMS devices.

To practically use the nanomotors for applications, it is important to evaluate the controllability, robustness, and efficiency in assembling and rotating arrays of nanomotors. We patterned and synchronously rotated ordered 2×2 and 1×3 arrays of nanowire motors as shown in Fig. 4(a) – (b) and Videos S2 – S3. Although the rotation speeds of the nanomotors were slightly different depending on the dimensions of individual nanowires and bearings, all nanomotors instantaneously rotated from the same initial positions with controlled rotation speed and chirality. They can instantly rotate, stop, and reverse the rotation orientations. The speed and chirality were precisely controlled by the voltages and phase shifts of the applied AC *E* field [Fig. 4(c)]. The nanomotors can robustly rotate for at least an hour.

How fast can the nanomotors ultimately rotate? To test the limit, we optimized the AC *E* field intensity and frequency. As aforediscussed, the rotation speed of nanomotors increases with $E^2$. Simply narrowing the quadrupole electrodes from a gap distance of 500 μm to 100 μm, we readily increased $E^2$ by 625 times. The rotation speed also depends on the applied AC frequency. For 10 μm Au/Ni/Au nanowires in DI water, the maximum rotation speed was obtained at 30 kHz [Fig. S8]. Applying these optimized conditions, we rotated nanomotors to a speed of at least 18000 rpm at 17 V and 30 kHz as shown in video S4 (1200 fps) and the inset of Fig. 5(a) and (b). The rotation orientation was clockwise as determined from the slowed motion in Video S5 (14 V). Agreeing with our other results, the rotation speeds (to at least 18000 rpm) were proportional to $V^2$, which counts for the ultrahigh speed rotation at only 17 V [Fig. 5(a)]. As far as we know, such a rotation speed is the highest achieved in natural or man-made motors of the same scale. It is of the same speed level of jet engine, but is still not the limit.

Finally, the applications of nanowire motors were demonstrated for rotation-controlled biomolecule release as shown in Fig. 6(a). By functionalizing the surface of nanowire rotors with surface-enhanced-Raman-scattering sensitive Ag nanoparticles,[37] we detected time-dependent release of biochemicals from single rotating nanomotors using Raman spectroscopy (video S10 and Figure S12). The release rate of biomolecules $k$ monotonically increases with the rotation speeds of nanomotors [Fig. 6(b) and supporting information]. For the first time, the release rate of drugs from nanoparticles can be precisely controlled in a mechanical manner. The detailed mechanism can be attributed to thickness change of electric-double layers, which will be studied elsewhere. This result demonstrates a pivotal application of nanomotors in controlled drug delivery, which is highly desirable for study of single cell stimulation, cell-cell communication, and system biology.

In summary, we demonstrated innovative design, assembly and actuation of rotary NEMS devices utilizing a strategic bottom-up nano-assembly and manipulation technique. Ordered arrays of nanomotors can be efficiently assembled and synchronously rotated with controlled angle, speed, and chirality. The rotation speed can be at least 18000 rpm, which is still not the limit. Various complex nanoscale forces and torques involved in the nanomotor system are quantitatively modeled and determined, which is critical for understanding, design, and actuation of various metallic NEMS devices. The nanowire motors were applied for drug release, and for the first time demonstrated that releasing rate of biochemicals from nanoparticles can be tuned by mechanical rotation. The multilevel innovations reported in this research may bring transformative impact to fields including NEMS, bioNEMS, microfluidics, and lab-on-a-chip architectures.

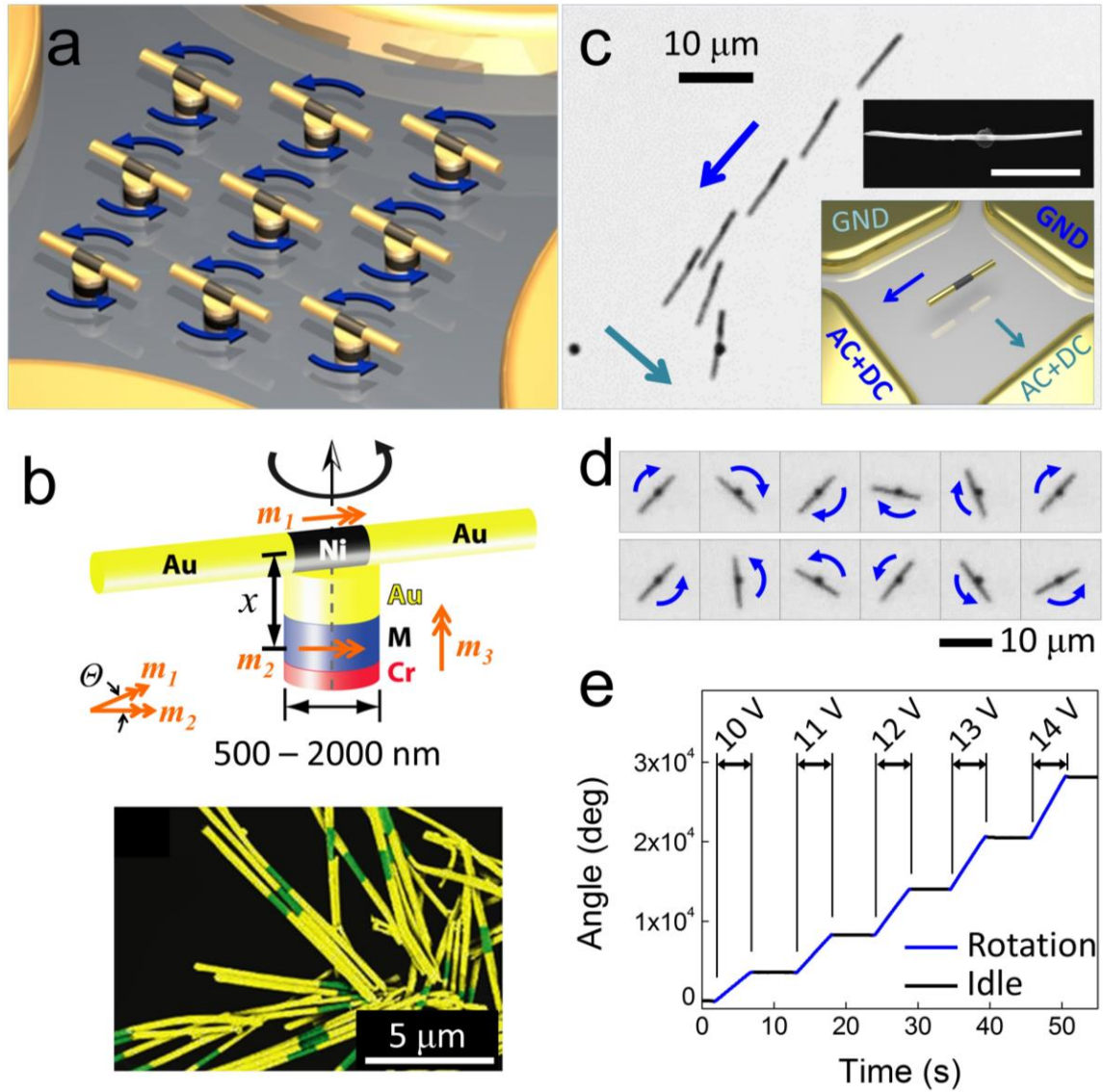

**Figure 1** (a) Schematic diagram of an array of nanomotors assembled from nanoscale building blocks with multi-segment Au/Ni/Au nanowires as rotators, tri-layer Au/Ni/Cr magnets as bearings, and microelectrodes as stators. (b) Schematic diagram of a nanomotor and a SEM image of Au/Ni/Au nanowires highlighted in false colors (yellow: Au; green: Ni). (c) Overlapped images of an Au/Ni/Au nanowire transported and assembled on a nanobearing by the electric tweezers. (Inset) SEM image of an assembled nanomotor (scale bar: 5 µm) and a schematic diagram of electric tweezers manipulating a nanowire. (d) Snapshots of a nanowire rotating on a nanobearing CW and CCW every 71 ms. (e) Angular position $\theta$ of an assembled nanowire in AC ***E*** fields of 10 – 14 V [rotor: Au(4.5 µm)/Ni(1 µm)/Au(4.5 µm); bearing: a thin film stack made of Au(100 nm)/Ni(80 nm)/Cr(6 nm) and 1 µm in diameter].

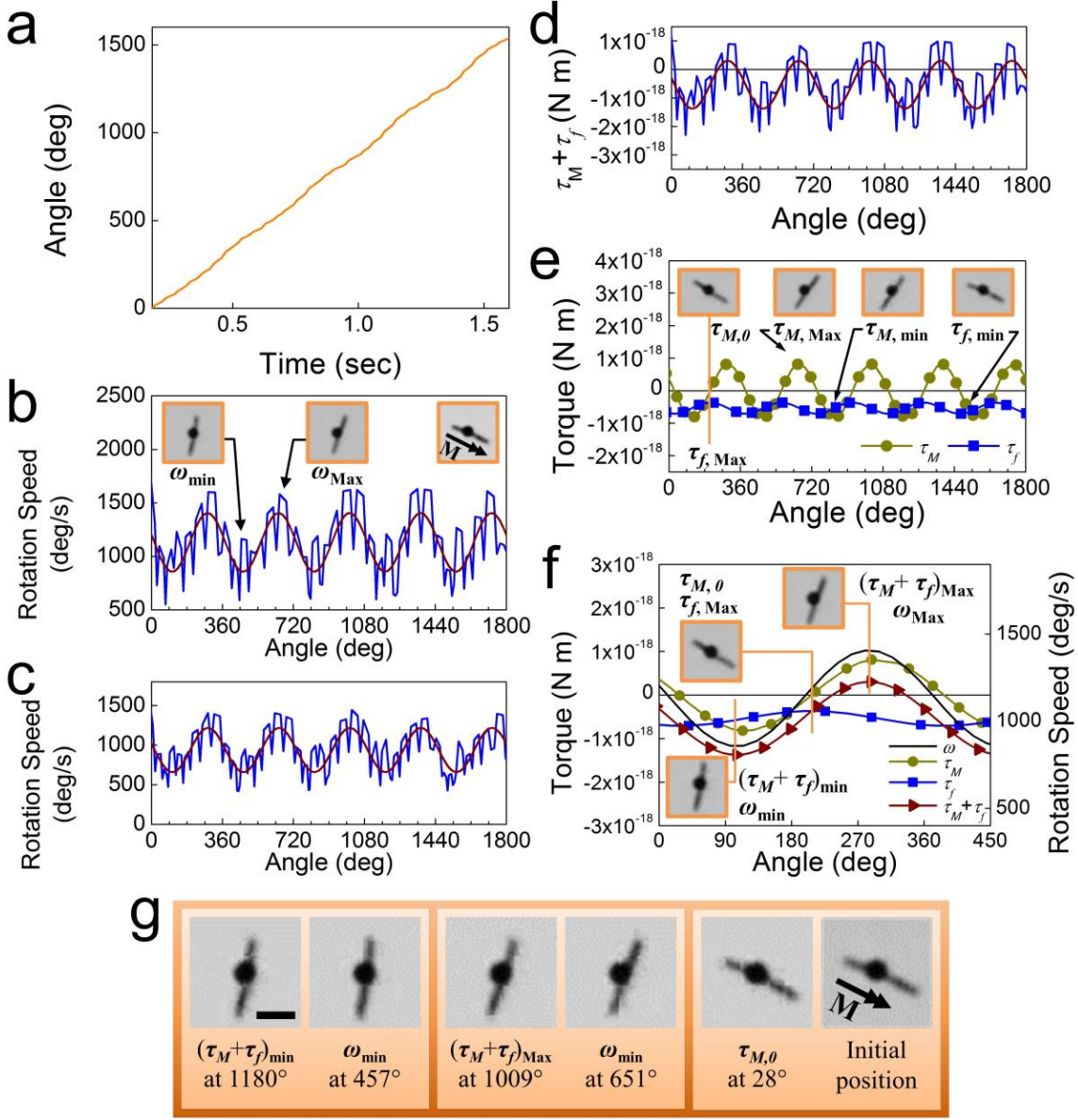

**Figure 2** (a) Rotation angle ($\theta$) versus time ($t$). (b – c) Rotation speed ($\omega$) versus angular position ($\theta$) of the nanomotor at (b) 12V and (c) 11 V, 10 kHz (CW) (d) Sum of the magnetic ($\tau_M$) and friction torques ($\tau_f$): ($\tau_M + \tau_f$) exerted on the motor at 12 V, 10 kHz. (e) Analytic solutions of $\tau_M$ and $\tau_f$ versus angle with the snapshots illustrating the angular position of the nanomotor for $\tau_{M,0}$, $\tau_{M,Max}$, $\tau_{M,min}$, $\tau_{f,Max}$, and $\tau_{f,min}$, (f) Angle-dependent torques of ($\tau_M + \tau_f$), $\tau_M$, $\tau_f$, and rotation speed $\omega$ of the nanorotor at 12 V, 10 kHz. The snapshots showed the angular positions of the nanomotor when the rotation speeds are the maximum ($\omega_{Max}$), minimum ($\omega_{min}$), the combined magnetic and frictional torque of ($\tau_M + \tau_f$) is maximum, ($\tau_M + \tau_f$)$_{min}$, and $\tau_{M,0}$, respectively. (g) The maximum/minimum rotation speeds of nanomotors occur when the combined torque of ($\tau_M + \tau_f$) is the maximum/minimum in the same direction of $E$ field. The alignment of the rotating nanomotor at $\tau_M = 0$ is the same as the initial alignment of the just-assembled nanowire. (Scale bar: 5 μm)

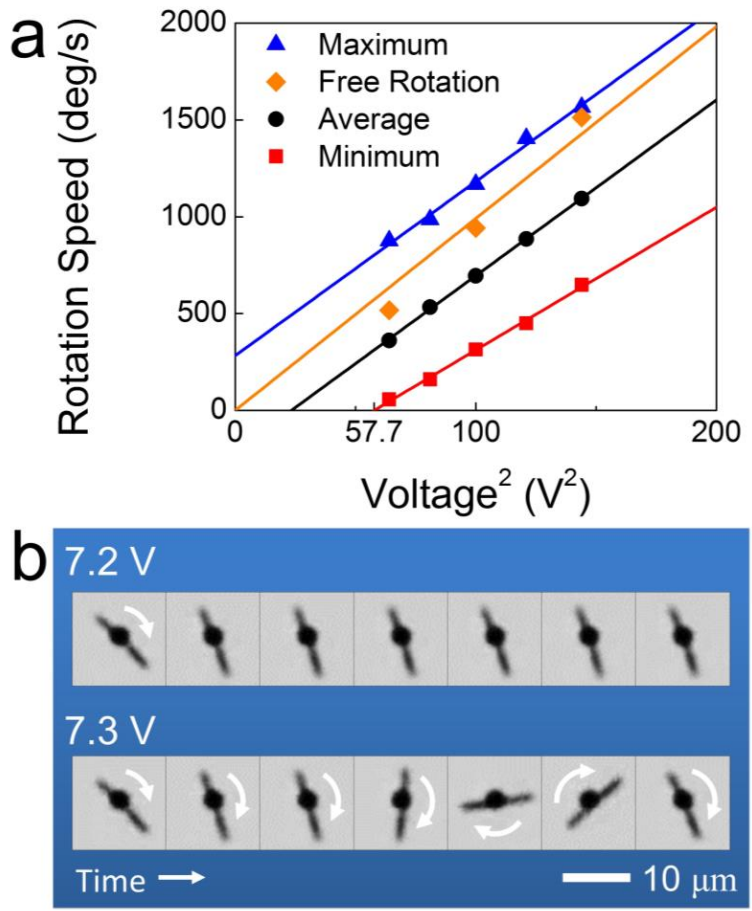

**Figure 3** (a) High (blue), low (red), and average (black) rotation speeds of a nanomotor (bearing diameter: 2 µm) linearly increase with $V^2$ (the same was observed for nanomotors with bearings of 500 nm- 1 µm as in the supporting information S9). For the same nanowire freely rotating in suspension, the rotation speeds increase with $V^2$ with a steeper slope (orange). (b) Snapshots of the nanomotor taken every 400 ms at AC voltages of 7.2 and 7.3 V, respectively, show the threshold voltage is 7.3 V.

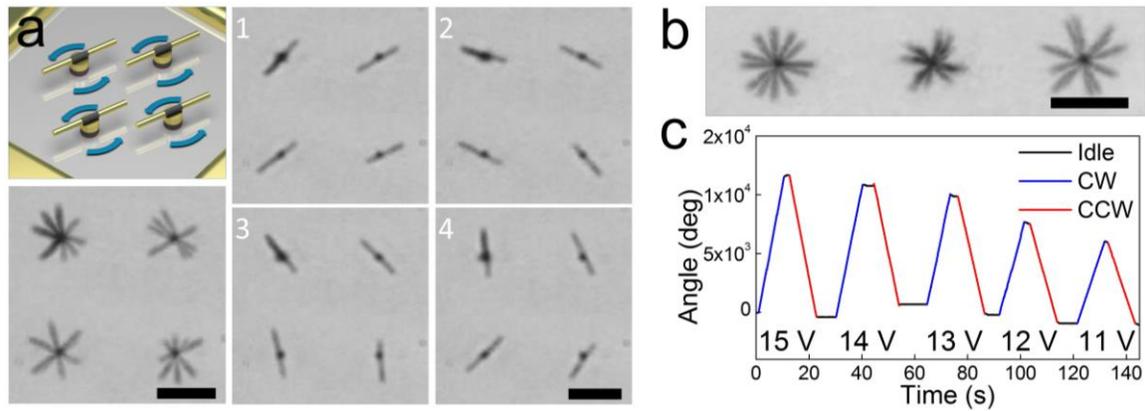

**Figure 2** (a) Schematic diagrams and snapshot images of a 2×2 nanomotor array, synchronous rotating clockwise at 12 V, 10 kHz. (b) Overlapped snapshot image of a 1×3 nanomotor array at 15 V, 10 kHz. The scale bars are 10 μm. (c) The rotation speed of nanomotors can be precisely controlled by AC $E$ fields as shown by rotation angle ($\theta$) versus time at 15 – 11 V, 10 kHz. The nanomotors can instantly rotate in orientations both CW (blue) and CCW (red) and stop on demand.

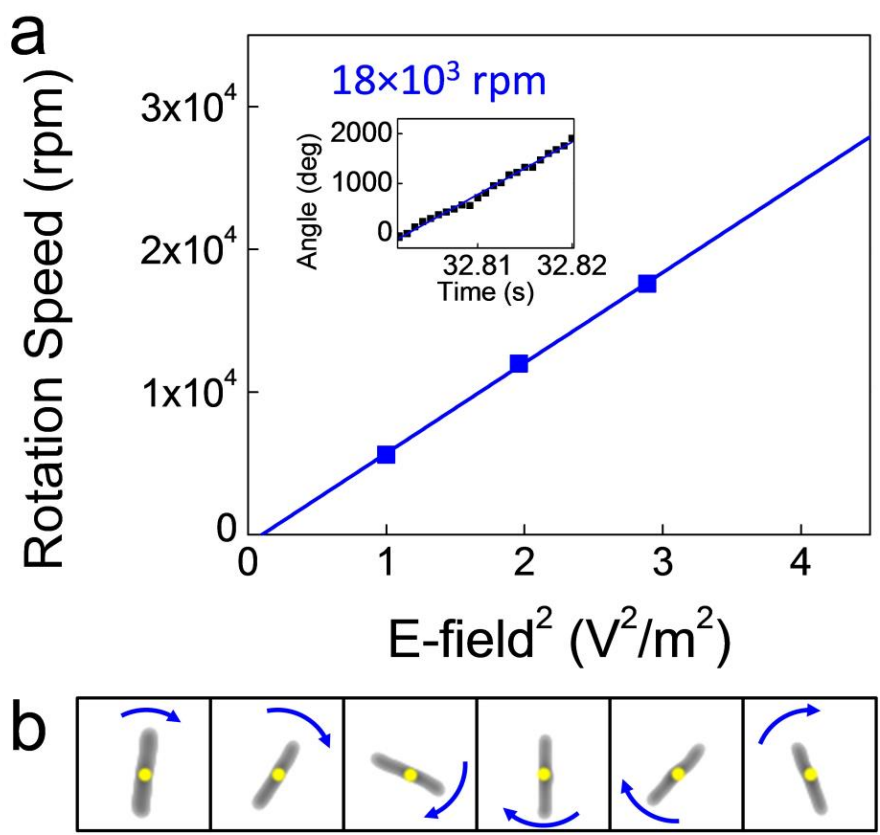

**Figure 3** (a) Rotation speed ($\omega$) of nanomotors from 10 – 17 V at 30 kHz in a 100 μm-gapped quadruple microelectrode. (Inset) Angle ($\theta$) versus time of the nanomotor rotating clockwise with a speed of ~18,000 rpm. (b) Enhanced snapshot images taken every 0.8 ms of the same nanomotor rotating at ~ 18,000 rpm at 17 V, 30 kHz.

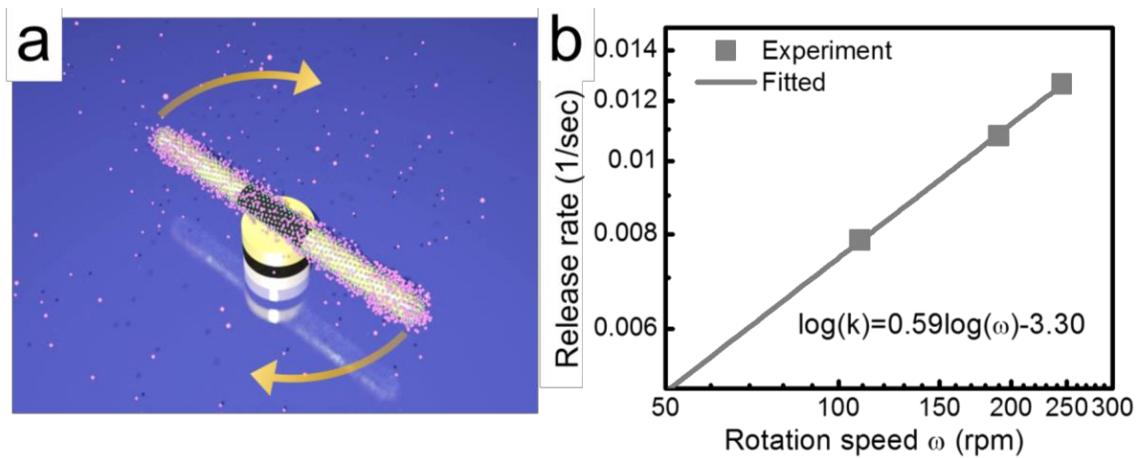

**Figure 6** (a) Illustration of rotation controlled biomolecule release (the motor fixed on the nanomagnet is SERS sensitive, which is a silica shelled Au-Ni-Au nanowire with uniform surface distributed Ag nanoparticles). The nanomotor is functionalized with biomolecules (in pink). (b) The release rate $k$ of biomolecules monotonically increases with the rotation speeds of nanomotors.

**Acknowledgement**

We are grateful for the support of National Science Foundation CAREER Award (Grant No. CMMI 1150767), Welch Foundation (Grant No. F-1734), and startup funds from the Cockrell School of Engineering of UT-Austin.

**Methods**

Fabrication of nanowires

In a three-electrode setup, Cu layer on the back of nanoporous AAO template, Pt mesh, and a Ag/AgCl electrode serves as a working electrode, a counter electrode, and a reference electrode, respectivly. The growth of the nanowires commences at the bottom of nanopores at the working electrode. The amount of electric charges passing through the circuit controls the length of each segment. As a result, arrays of nanowires, with a structure of 1-μm-long Ni segment sandwiched between two 4.5-μm-long Au segments, were synthesized and suspended in DI water.

Creation of rotating $E$ fields

Rotating AC $E$ fields can be generated in the center of quadruple microelectrodes by applying four AC voltages with 90° sequential phase shifts on the four sub-electrodes.[38]

Supporting Information

# Ultrahigh-Speed Rotating Nanoelectromechanical System (NEMS) Devices Assembled from Nanoscale Building Blocks

Kwanoh Kim[1], X. B. Xu[2] and D. L. Fan[1, 2]

[1] Department of Mechanical Engineering, the University of Texas at Austin, Austin, TX 78712, USA
[2] Materials Science and Engineering Program, the University of Texas at Austin, Austin, TX 78712, USA
Correspondence can be addressed to: dfan@austin.utexas.edu

**Video S1:** The Au/Ni/Au nanowire (10-μm in length and 300 nm in diameter) was precisely assembled on the tri-layer nanobearing by using the electric-tweezer manipulation technique. The magnetic attraction between the nanowire and the nanobearing anchored the nanowire on the nanobearing, while still allowed its rotation. The video was enhanced.

**Video S2:** Synchronous rotation of 1×3 arrays of nanomotors at 15 – 13 V and 10 kHz.

**Video S3:** Synchronous rotation of 2×2 arrays of nanomotors at 12 V and 10 kHz.

**Video S4 and S5:** Ultrahigh-speed rotation of nanomotors to at least 18000 rpm at 17 V and 30 kHz. The enhanced video S4 shows the real time motion of the nanowire rotor at 17 V. The enhanced video S5 shows the rotation of the same nanomotor slowed by 40 times at 14 V, confirming the rotation orientation is clockwise.

**S6: Magnetic alignment of a nanowire assembled on a nanobearing with perpendicular magnetic anisotropy (PMA)**

The nanobearing with the Ni magnetic layer deposited at a rate > 1 Å/s has a perpendicular magnetic anisotropy (PMA) due to its stress induced magneto-restriction.[1] The Ni layer uniquely has a slanted PMA which consists of a bigger perpendicular component ($m_3$) and a smaller in-plane component ($m_2$) [Fig. S6(a)], where $m_3$ can be much smaller after treatment with an in-plane oscillating magnetic field. The magnetic hysteresis loops obtained by a vibrating sample magnetometer (VSM) show the strong perpendicular magnetic anisotropy of Ni [Fig. S6(b)]. The in-plane magnetic component ($m_2$) can be readily known from the interactions of the nanowire (magnetic moment of

$m_1$) with the nanobearing, which result in spontaneous alignment of the nanowire in the magnetic treatment direction of the bearing as shown in video S1.

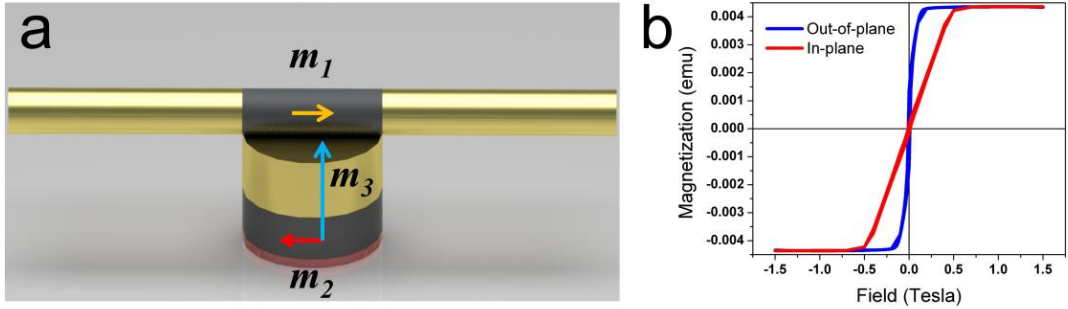

**Figure S6** (a) A Au/Ni/Au nanowire with a magnetic moment of $m_1$ is assembled on a nanobearing consisting of a thin film stack of Au/Ni/Cr. The Ni layer in the nanobearing has a slanted perpendicular magnetic anisotropy (PMA) which consists of a perpendicular component ($m_3$) and an in-plane component ($m_2$). After treatment with an in-plane oscillating magnetic field, $m_3$ can be much smaller. The nanowire spontaneously aligns due to the magnetic interaction between $m_1$ and $m_2$ and its orientation is determined by the treatment direction of the external magnetic field. (b) VSM magnetic hysteresis loops of a 200 nm Ni thin film show the PMA nature of the Ni magnetic bearing.

**S7: Magnetic and friction torques at different AC voltages**

Using the same approach in extracting the magnetic and friction torques for the nanorotor rotating at 12 V, we obtained the angular dependent magnetic torque at other voltages, e.g. 10 and 11V as follows, which excellently support the feasibility of our modeling, where the magnetic and frictional torques in nanomotors are independent of the applied AC voltages (***E***-fields):

At 10 V:

$\tau_M = -7.33 \times 10^{-19} \sin(\theta - 26°)$

$\tau_f = -1.44 \times 10^{-19} \cos(\theta - 26°) - 5.41 \times 10^{-19}$

At 11 V:

$\tau_M = -8.36 \times 10^{-19} \sin(\theta - 26°)$

$$\tau_f = -1.39 \times 10^{-19} \cos(\theta - 26°) - 4.84 \times 10^{-19}$$

**S8: Optimization of the frequency of the AC *E* fields**

The rotation of nanowire nanomotors in DI water depends on both the dimensions and materials of the nanowires, as well as the frequency of the AC *E* fields. Frequency dependent rotation of 10-µm-long Au/Ni/Au nanowire rotors were investigated at from 1 – 150 kHz. The highest rotation speed was obtained at 30 kHz as shown in Fig. S8, at which we achieved ultrahigh-speed rotation of nanomotors to 18000 rpm.

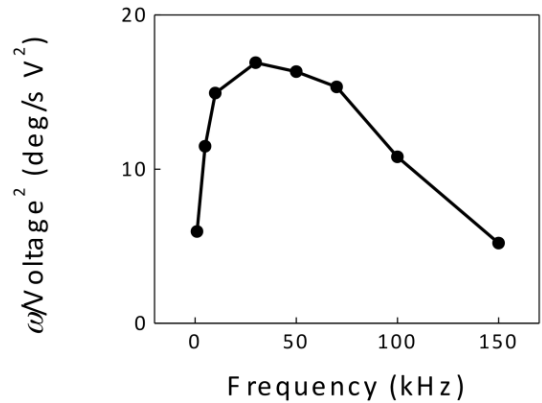

**Figure S8** ω/V² versus AC frequency for free rotation of 10-µm-long Au/Ni/Au nanowires . The highest rotation speed is obtained at 30 kHz.

**S9: Rotation of nanomotors with nanobearings of 1 µm to 500 nm in diameter**

The feature size of the nanomotors can be further scaled down by reducing the size of nanobearing from 2 µm to 1 µm [Fig. S9(a)] and 500 nm in diameter [Fig. S9(b)]. The dynamic characteristics of the nanomotors were analyzed with the same methods discussed in the main text, which shows similar linear relationship of *ω*-Voltage² with that found in nanomotors with bearings of 2 µm in diameter [Fig. 3(a)].

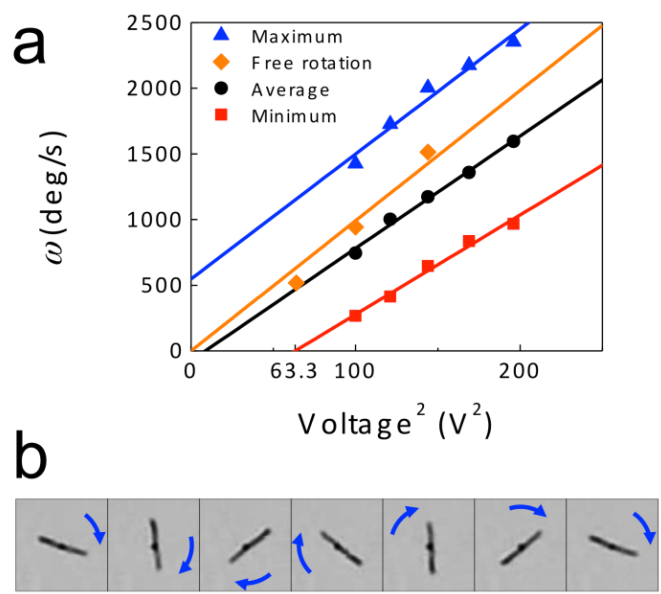

**Figure S9** (a) Rotation of nanomotors with magnetic bearings with sizes from 500 nm to 1 μm. The high (blue), low (red), and average (black) speeds of a rotating nanorotor [Au(4.5 μm)/Ni(1 μm)/Au(4.5 μm)] on an 1-μm magnetic bearing (Au(100 nm)/Ni(80 nm)/Cr(6 nm). (b) Snap shots of a nanomotor rotating on a 500-nm magnetic bearing every 67 msec.

**Video S10:** Raman video of biochemical molecules on a rotating SERS-sensitive nanomotor

**S11: Fabrication of nanowire rotors with surface-distributed-Raman-sensitive Ag nanoparticles**

Au-Ni-Au nanowires were firstly coated with a layer of 70 nm silica via hydrolysis of tetraethyl orthosilicate and then Ag nanoparticles were grown on the surface of the silica by a wet chemistry method.[2] The Raman-sensitive nanomotors fabricated by this method can provide an Raman enhancement factor of $10^{10}$ with estimated Ag NP density of 1600/μm$^2$ and hotspot density of 1200/μm$^2$.[2] The detailed silica/Ag nanoparticles coating process can be found in our previous work.[2]

**S12: Time-dependent biomolecule release controlled by rotation of nanomotors**

We functionalized biochemicals such as nile blue (NB) on Raman-sensitive nanowire rotors by incubating the nanowires in NB solutions (340 nM) for 30 min. Then we removed excess nile blue on the nanowires with abundant D.I. water and assembled nanowires into nanomotors by electric-tweezer manipulation. The NB molecules on nanomotors were detected from their surface enhanced Raman scattering (SERS) spectra, which were excited by a 532 nm laser and collected by a high-sensitive spectrograph and CCD camera. In this way, the time-dependent biomolecule release can be readily recorded in real time from a single rotating nanomotor.

The correlation of NB concentration and SERS intensity was obtained for concentrations of 5 nM to 340 nM. Simply modeling the rotation-controlled molecular release as a gradient induced diffusion process, we can calculate the time-dependent molecule release from the Fick's first law, given by:

$C = C_1 e^{-kt} + C_0,$

where $C$ is the concentration of molecules on the surface of the nanomotor at time $t$, $C_1$ is the initial concentration of molecules on the nanomotors when $t=0$, $C_0$ is the concentration of molecules in the bulk solution, $k$ is the molecule release rate (unit 1/sec), which depends the nanomotor geometry, thickness of the silica layer, molecular species, and the thickness of electric double layer between the surface of nanomotor and bulk solution. The molecule release rate ($k$) indeed can be systematically changed by the rotation speed of nanomotors ($\omega$) [Fig. S12(a)]. We attributed it to the tuning of the thickness of electric double layer by mechanical rotation of nanomotors (to be discussed in the future). This work, for the first time, demonstrated a new mechanism for controllable drug release [Fig. 6(b)].

To confirm that the change of molecule release rate by rotating nanomotors is not a result of the AC electric field, we performed a control experiment by turning AC fields on/off to a non-moving nanorotor attached to the substrate while detecting molecule release at the same time. Clearly the AC electric field cannot alter the molecule release rate as shown in Fig. S12(b).

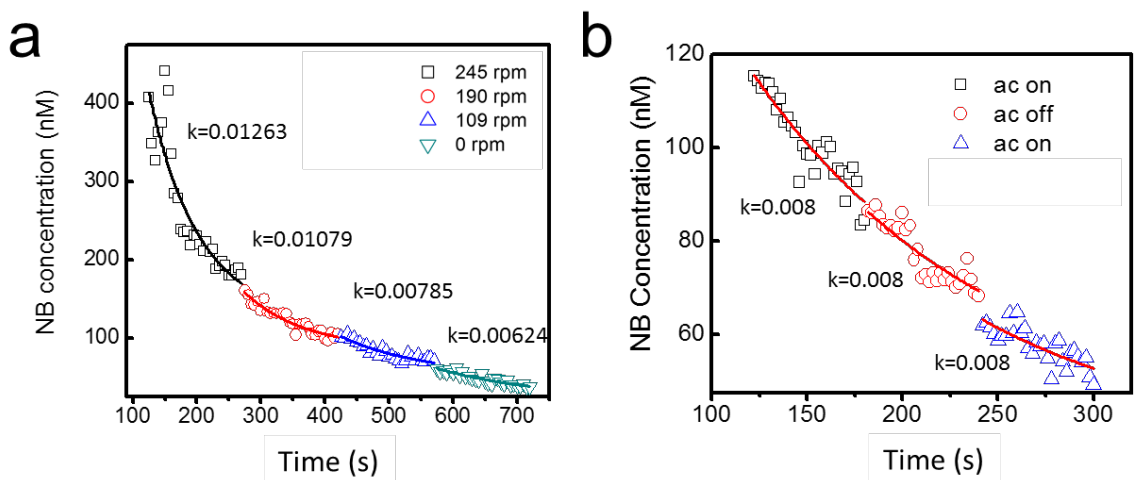

**Figure S12** (a) Rotation speed controlled time-dependent molecule release. (b) Control experiment: the AC electric field cannot alter the molecule release rate.